\documentclass[%
 aip,
 amsmath,amssymb,
 reprint,%
]{revtex4-1}
\usepackage{color}
\usepackage{comment}
\usepackage{ulem}

\usepackage{graphicx}% Include figure files
\usepackage{dcolumn}% Align table columns on decimal point
\usepackage{bm}% bold math
\usepackage[utf8]{inputenc}
\usepackage[T1]{fontenc}
\usepackage{mathptmx}
\usepackage{etoolbox}

\makeatletter
\def\@email#1#2{%
 \endgroup
 \patchcmd{\titleblock@produce}
  {\frontmatter@RRAPformat}
  {\frontmatter@RRAPformat{\produce@RRAP{*#1\href{mailto:#2}{#2}}}\frontmatter@RRAPformat}
  {}{}
}%
\makeatother

\begin{document}

\preprint{AIP/123-QED}

\title[]{Hydrogenation-induced gigantic resistance decrease of palladium films deposited by high pressure magnetron sputtering}

\author{Yusuke Ikeda}
\affiliation{Department of Basic Science, University of Tokyo, Meguro, Tokyo 153-8902, Japan}

\author{Takuya Kawada}
\affiliation{Department of Basic Science, University of Tokyo, Meguro, Tokyo 153-8902, Japan}
\email{takuyakawada@g.ecc.u-tokyo.ac.jp}

\author{Yuki Shiomi}
\affiliation{Department of Basic Science, University of Tokyo, Meguro, Tokyo 153-8902, Japan}
%\email{yukishiomi@g.ecc.u-tokyo.ac.jp}

\date{\today}

%TC:ignore
\begin{abstract}

%Hydrogenation of palladium (Pd) causes a change in electrical resistance, which can be employed for resistive hydrogen sensors. Achieving larger resistance changes is desirable for enhancing device sensitivity; however, the fabrication of Pd films exhibiting substantial resistance variations typically requires sophisticated techniques. In this work, 
We demonstrate a pronounced decrease in the electrical resistance of highly disordered palladium (Pd) films deposited under a high working Ar pressure using a compact film coating system. The resulting resistance change ratio of up to $1/335$ is predominant among those reported previously. Film characterization suggests two primary mechanisms responsible for this significant resistance reduction: atomic force microscopy observation indicates improved electrical contacts among Pd grains, and X-ray diffraction measurement demonstrates hydrogenation-induced crystallization of Pd. These findings offer a simple scheme to enhance hydrogen sensor performance and can contribute to a more comprehensive understanding of the hydrogenation process in Pd.

\end{abstract}
%TC:endignore

\maketitle

%\begin{quotation}

%\end{quotation}

Hydrogen is an energy carrier that can be generated from renewable energy sources and is regarded as a key component of decarbonization~\cite{fang2024acs}. A stable supply of hydrogen-based clean energy requires efficient storage and transportation systems. To this end, a variety of hydrogen storage materials have been explored~\cite{kalibek2024}. Among them, palladium (Pd) has been extensively investigated as a prototypical material owing to its ability to reversibly absorb large amounts of hydrogen, even under ambient temperature and pressure conditions~\cite{kay1986prb}, making it a promising candidate for practical hydrogen storage~\cite{brian2011,dekura2019}.

Hydrogen absorption by Pd substantially modifies its optical, electrical, magnetic, and spintronic properties~\cite{lin2013,chang2019,skoskiewicz1973,fischer1906,masanori2006,ogata2022}. These property changes enable diverse applications, including monitoring the hydrogenation state of Pd, hydrogen detection, and realization of hydrogen-tunable Pd-based functional devices, thereby contributing to the safe and efficient utilization of hydrogen energy. Here, we focus on the modulation of the electrical properties induced by hydrogenation. Numerous studies have demonstrated that the resistance of Pd is affected by its hydrogen absorption. The mechanism underlying this resistance modulation is intrinsically linked to the details of the hydrogen absorption process, which involves molecular adsorption, dissociation into atomic hydrogen, and subsequent diffusion into the Pd lattice. Each of these steps is strongly influenced by parameters such as film thickness, grain size, substrate type, and surface pretreatment~\cite{fromm1987,duncan2008,flanagan2005}, leading to considerable tunability in both the polarity and magnitude of resistance changes.

In bulk Pd, resistivity typically increases %by at most a factor of two at room temperature, 
because of Pd hydride formation and/or enhanced electron scattering due to the absorbed hydrogen atoms~\cite{YSakamoto_1996,geerken1983}. Conversely, resistivity decreases have also been reported. For example, in discontinuous Pd ultrathin films and mesowires, hydrogen-induced volume expansion enhances interparticle electrical contact, thereby lowering resistivity~\cite{xu2005,frederic2001,singh2020}. In addition, a reduction in resistivity has been observed in Pd wires exposed to hydrogen at elevated temperatures, possibly owing to hydrogenation-induced recrystallization~\cite{xu1998}. 
In practical scenarios, multiple mechanisms often act simultaneously, leading to both positive and negative resistance responses, depending on the specific material conditions~\cite{fan1994,das2020}.

\begin{figure}[t]
    \includegraphics[width=1\linewidth]{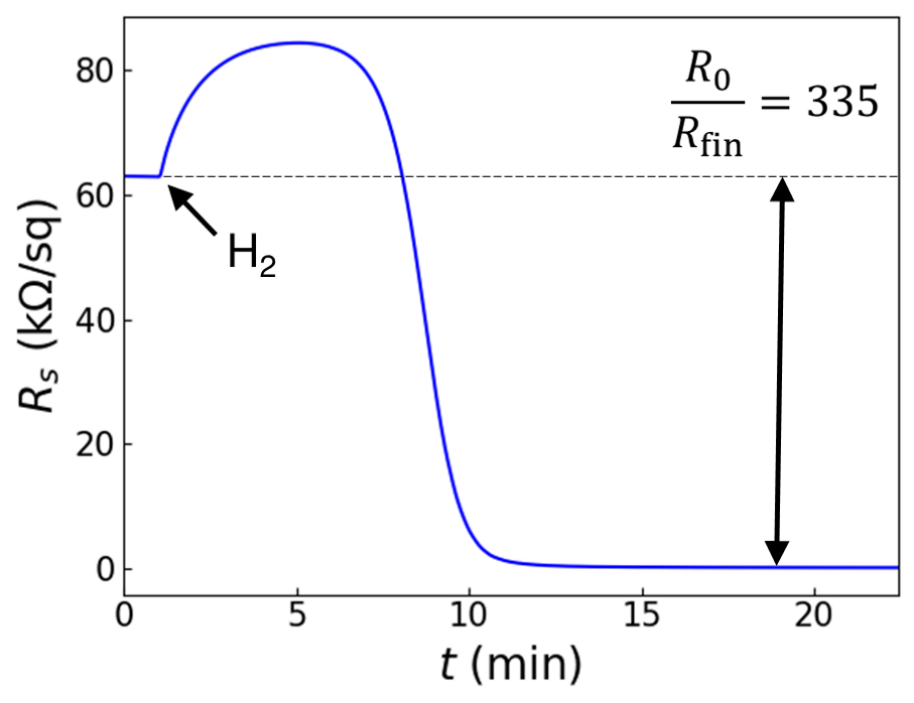}
    \caption{\label{Champion} Sheet resistance ($R_\mathrm{S}$) as a function of elapsed time for Pd(7.5 Pa). The dotted horizontal line indicates the value of initial resistance ($R_0$). The resitance change ratio with respect to $R_{\rm fin}= 188$ $\Omega$/sq is also written. }
\end{figure}

The change in resistance induced by hydrogenation provides a basis for resistive hydrogen sensors using Pd thin films. Although improving sensor sensitivity requires maximizing this resistance change, Pd thin films grown by conventional deposition methods usually exhibit resistance changes of a few \% to 20 \%~\cite{OZTURK2016179} at most.
On the other hand, previous efforts succeeded in enhancing resistance change upon hydrogen absorption, reaching factors of up to 100 in some cases ~\cite{vanlith2007,lim2012}. However, the fabrication of Pd films exhibiting such large resistance variations typically relies on sophisticated techniques, including cluster deposition~\cite{reichel2006,vanlith2007} and multistep fabrication processes for Pd nanotubes~\cite{lim2012}.

\begin{table*}[bht]
    \caption{\label{tab:devices}
    Summary of Pd deposition conditions and representative film characteristics.
    }
    \begin{ruledtabular}
        \begin{tabular}{cccccc}
             ID & $P_{\rm Ar}$ [Pa] & $t_{\rm depo}$ [s] & $d_{0}$ [nm] & $d_{\rm fin}$ [nm] & $\rho_{0}$ [$\mu \Omega \cdot $cm] \\ \hline
             Pd(2.5 Pa) & 2.5 &  75 &  60 &  61 &  26.2  \\ 
                Pd(5.0 Pa) & 5.0 & 150 &  95 &  92 &   186  \\
                Pd(7.5 Pa) & 7.5 & 225 & 178 & 181 & $6.62 \times 10^{5}$ \\
                Pd(10 Pa)  &  10 & 300 & 225 & 200 & $2.61 \times 10^{3}$ \\
                Pd(15 Pa)  &  15 & 450 &  93 &  94 &   624 \\
                Pd(20 Pa)  &  20 & 600 & 114 & 109 &   738 \\
        \end{tabular}
    \end{ruledtabular}
\end{table*}

Our strategy to enhance the resistance decrease by hydrogenation is based on previous studies that reported large resistance decrease for disconnected Pd films with a high resistivity~\cite{vanlith2007,das2020}. We deduced that disordered films containing numerous defects and/or vacancies, amorphous structures, and pronounced surface roughness are likely to exhibit significant resistance changes.

In this study, we demonstrate a simple method for preparing Pd thin films that exhibit significant resistance changes upon hydrogen absorption. The films were deposited on glass substrates by DC magnetron sputtering under a high working argon (Ar) pressure. Consequently, resistance changes of up to a factor of $< 1/300$ were achieved, exceeding the values reported to date. The properties of the films were examined to elucidate the underlying mechanisms. In some Pd films, the resistance change ratio correlated with the initial resistivity, suggesting that resistance modulation is predominantly caused by the improvement of electrical contacts among Pd particles. This is supported by atomic force microscopy (AFM), which observed a granular structure on such film surfaces and an increase in the grain size after the hydrogen exposure. 
In contrast, the other films exhibited large resistance changes, despite the relatively low initial resistivity. In such films, X-ray diffraction (XRD) analysis revealed the emergence of a Pd(111) diffraction peak after hydrogen exposure whereas the as-deposited films were amorphous, indicating that hydrogen-induced crystallization is the dominant mechanism. These results offer a simple and effective approach for enhancing hydrogen sensor performance and provide new insights into the hydrogenation process in Pd.

Pd films were deposited on glass substrates by DC magnetron sputtering using a compact film coater equipped with a rotary pump (SC-701 Mk II ADVANCE, Sanyu Electron). The Pd target has a purity of 99.9\% and a diameter of 50 mm. The sputtering conditions are summarized in Tab.~\ref{tab:devices}. In the sputtering method, the Ar gas pressure is usually kept low in order to improve film quality. To obtain highly resistive films, however, we kept the working Ar gas pressure ($P_{\rm Ar}$) high, above 2 Pa, during deposition. Hereafter, the film deposited at $P_{\rm Ar}=x$ Pa is referred to as Pd($x$ Pa). Because a high Ar pressure reduces the mean free path of the sputtered atoms, the deposition time $t_{\rm depo}$ was increased accordingly for a higher $P_{\rm Ar}$, as shown in Tab.~\ref{tab:devices}. The base pressure prior to deposition was 1.6 Pa. The deposition was performed at room temperature, and the target--substrate distance was fixed at 30 mm. The deposited films were patterned into a Hall bar geometry with dimensions of 1.3 mm (length) $\times$ 1.0 mm (width) using a metal mask.

The resistance response of Pd films to hydrogen exposure was investigated using a four-terminal configuration. To enhance the signal-to-noise ratio, resistance measurements were carried out using a lock-in technique: an AC current of 1 mA at 78 Hz was applied to the film by a source meter (6221 DC and AC current source, Keithley), and the resulting AC voltage was monitored using a lock-in amplifier (LI5655, NF Corporation). 
%For some films, we confirmed that the similar results were obtained for 1 $\mu$A: see Figs.~S1 and S2 in the supplementary material. 
During the measurements, the films were enclosed in a sealed steel chamber connected to a rotary pump and balloon containing a 3 vol\% H$_2$/97 vol\% Ar gas mixture via manual valves. The chamber was evacuated using the pump before the introduction of hydrogen. The H$_2$/Ar mixture was then introduced 60 s after the start of the resistance recording, and the measurements continued until the resistance variation became negligible.

\begin{figure}[tb]
    \centering
    \includegraphics[width=1\linewidth]{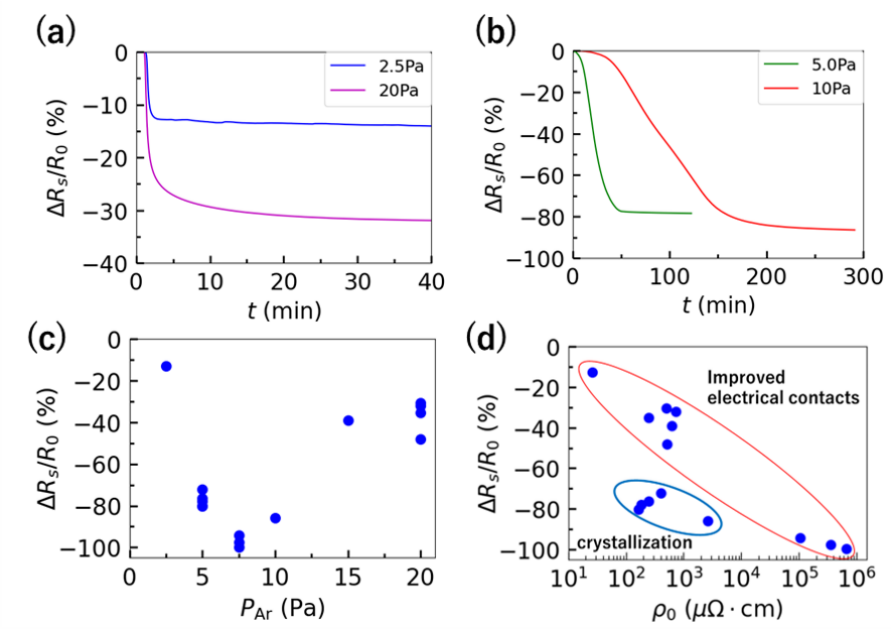}
    \caption{(a),(b) Sheet resistance change ratio ($\Delta R_\mathrm{S}$/$R_0 \equiv (R_s -R_0)/R_0 $) as a function of elapsed time for Pd(2.5 Pa) and Pd(20 Pa) (a), and for Pd(5.0 Pa) and Pd(10 Pa) (b). (c) Working Ar pressure ($P_{\rm Ar}$) dependence of $\Delta R_\mathrm{S}/R_0$. (d) $\Delta R_\mathrm{S}/R_0$ plotted against initial resistivity ($\rho_0$). We classified the data into two groups (see text).}
    \label{resistance}
\end{figure}

Figure~\ref{Champion} presents the representative data obtained in this study. The sheet resistance ($R_\mathrm{S}$) of Pd(7.5 Pa) is plotted as a function of the elapsed time after the start of the measurement. The resistance significantly changed by hydrogen exposure, and almost saturated in about 10 min. The initial and final sheet resistances were 62.9 k$\Omega$/sq and 188 $\Omega$/sq, respectively, corresponding to a resistance change ratio of 335. This value is comparable to the highest values ever reported~\cite{vanlith2007}. The time-series data reveals that $R_\mathrm{S}$ initially increases immediately following hydrogen exposure, reaches a maximum, and subsequently decreases markedly. 
The initial increase in resistance can be attributed to enhanced carrier scattering due to absorbed hydrogen atoms~\cite{YSakamoto_1996,geerken1983,lee2010,zeng2012}.
%: see the supplementary material for the detail.
%Based on the scenarios proposed in previous studies, the initial increase in resistance can be attributed to Pd hydride formation and/or enhanced carrier scattering by absorbed hydrogen atoms.
The subsequent decrease in resistance is commonly ascribed to improvement of electrical contacts among Pd grains and hydrogenation-induced crystallization~\cite{xu1998,xu2005,frederic2001,singh2020}, although the magnitude of the change observed here is notably large.

To elucidate the mechanisms responsible for the gigantic resistance decrease, we characterized Pd films prepared at different $P_{\rm Ar}$ values ranging from 2.5 Pa to 20 Pa and measured their resistance response to hydrogen exposure. The film properties were evaluated both before and after hydrogenation. The film thickness was determined using a stylus profiler (Dektak, Bruker). Resistance measurements were performed using the same method as that for Pd(7.5 Pa). The initial thickness ($d_0$), final thickness after hydrogenation ($d_{\rm fin}$), and initial resistivity ($\rho_0$) of the films are listed in Tab.~\ref{tab:devices}. For Pd(5.0 Pa), Pd(7.5 Pa), and Pd(20 Pa), the values for a representative film are presented. The change in the thickness after hydrogen exposure is a few \% for most samples, except for a relatively large change of 11\% for Pd(10 Pa). To obtain information on film structures, the crystallinity was assessed using X-ray diffraction (XRD) (SmartLab, Rigaku). In addition, we prepared another Pd(5.0 Pa) and Pd(20 Pa), whose surface morphologies were examined using AFM (Nanocute, SII NanoTechnology). 
%We note that the similar resistance changes by hydrogen exposure were observed for these films: see Fig.~S1 in the supplementary material. 

Figures~\ref{resistance}(a) and (b) show the time evolution of the sheet resistance for Pd(2.5 Pa), Pd(5.0 Pa), Pd(10 Pa), and Pd(20 Pa). Here, $R_0$ denotes the initial sheet resistance, and $\Delta R_\mathrm{S}$ is defined as $R_\mathrm{S} - R_0$. All the samples exhibit negative resistance change by hydrogen exposure. The results indicate that $R_0$, $\Delta R_\mathrm{S}/R_0$, and the characteristic timescale of the resistance change depend strongly on $P_{\rm Ar}$. Based on the time dependence, we categorized the data into two groups, as shown in Figs.~\ref{resistance}(a) and (b): one in which the resistance decreases rapidly within 10 min and the other in which the decrease proceeds more slowly. Pd(2.5 Pa) and Pd(20 Pa) belongs to the first group, while Pd(5.0 Pa) and Pd(10 Pa) to the second one. Additionally, as already shown in Fig.~\ref{Champion}, Pd(7.5 Pa) is categorized into the first group, as the resistance change is almost complete in 10 min.

To clarify the overall trends, we plot the sheet resistance change $\Delta R_\mathrm{S}/R_0$ as a function of the working Ar gas pressure $P_{\rm Ar}$ in Fig.~\ref{resistance}(c). For Pd(5.0 Pa), Pd(7.5 Pa), and Pd(20 Pa), several films were measured to verify reproducibility. The magnitude of $\Delta R_\mathrm{S}/R_0$ increases with $P_{\rm Ar}$ and exhibits a maximum (99.7\%) at $P_{\rm Ar}=7.5$ Pa, followed by the decrease as $P_{\rm Ar}$ further increases. This nonmonotonic trend suggests that increasing Ar pressure does not necessarily enhance the resistance change.

\begin{figure}[t]
    \centering
    \includegraphics[width=1\linewidth]{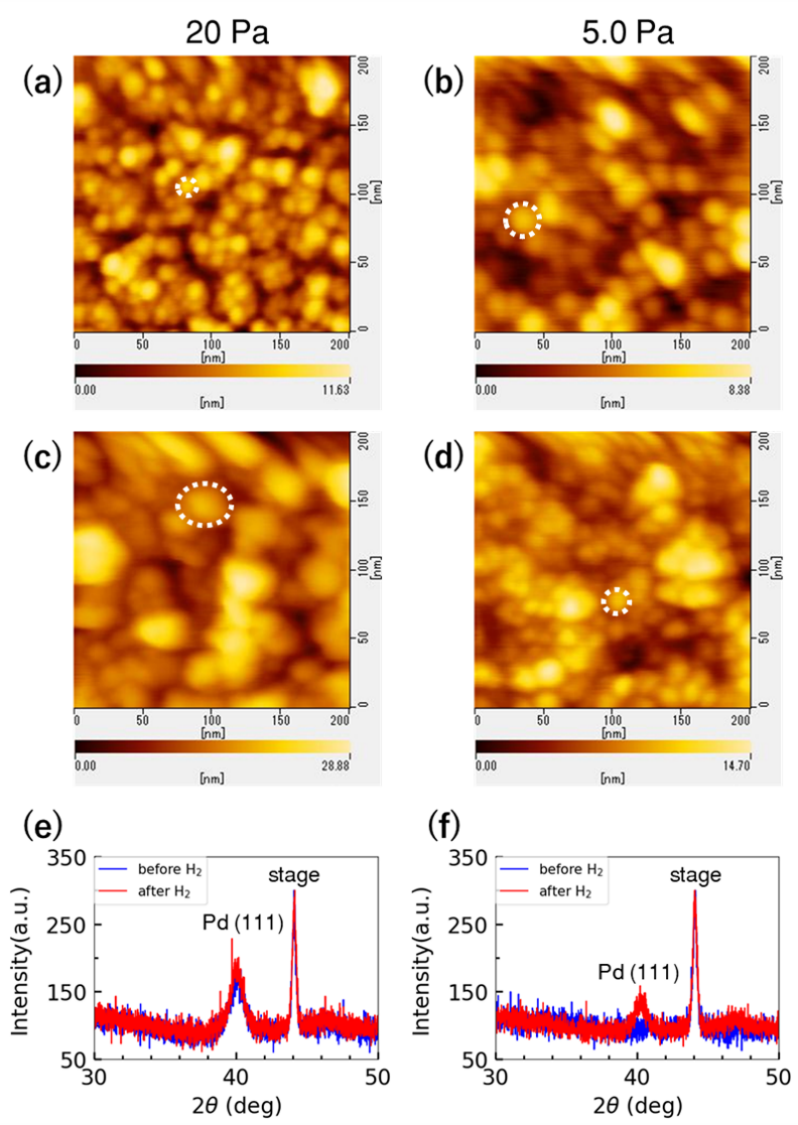}
    \caption{
    (a)-(d) AFM images of the surfaces of Pd(20 Pa) (a),(c) and Pd(5.0 Pa) (b),(d) taken before (a),(b) and after (c),(d) the hydrogen exposure, respectively. Grains with a typical size are indicated by white dotted circle, whose diameters are 13 nm (a), 22 nm (b), 33 nm (c), and 18 nm (d), respectively. (e),(f) XRD patterns of Pd(20 Pa) (e) and Pd(5.0 Pa) (f). Blue and red data represent the results obtained before and after the hydrogen exposure, respectively. Intensity is normalized by the peak from sample stage at $2\theta \sim 44^\circ$.
    %(a),(b) AFM images of the surfaces of Pd(5.0 Pa) (a) and Pd(20 Pa) (b). (c),(d) XRD patterns of Pd(5.0 Pa) (c) and Pd(20 Pa) (d). Blue and red data represent the results obtained before and after the hydrogen exposure, respectively.
    %(a)-(d) AFM images of the surfaces of Pd(5.0 Pa) (a),(c) and Pd(20 Pa) (b),(d) taken before (a),(b) and after (c),(d) the hydrogen exposure, respectively. (e),(f) XRD patterns of Pd(5.0 Pa) (e) and Pd(20 Pa) (f). Blue and red data represent the results obtained before and after the hydrogen exposure, respectively. Intensity is normalized by the peak from sample stage at $2\theta \sim 44^\circ$.
    }
    \label{afm}
\end{figure}

In Fig.~\ref{resistance}(d), we analyze the relationship between the sheet resistance change $\Delta R_\mathrm{S}/R_0$ and initial resistivity $\rho_0$. % (see Tab.~\ref{tab:devices} for the $\rho_0$ value of each sample). 
This plot demonstrates that films with a higher initial resistivity tend to exhibit larger resistance changes, which is consistent with earlier studies on disconnected Pd films~\cite{vanlith2007,das2020}. This behavior can be attributed to the improvement of the electrical contact among the grains associated with the hydrogenation, which is supported by AFM observations of the film surface.
%This behavior can be attributed to the expansion of Pd grains during hydrogen absorption, which improves the electrical contact between the grains. 
Figures~\ref{afm}(a) and (c) show the AFM images of Pd(20 Pa) before and after the hydrogen exposure, both of which reveal granular structures. Specifically, the grain size significantly increased after the hydrogen exposure, implying that the hydrogen absorption improved the electrical contact among Pd grains.
%On the other hand, no significant increase of the grain size was observed for Pd(5.0 Pa). This implies another factor associated with the resistance decrease in Pd(5.0 Pa), \sout{which is discussed later.} %The expansion of Pd grains is reasonable for hydrogen absorption and can be a cause of the improvement of electrical contacts. This phenomenon is prominent in the first group (see Fig.~\ref{resistance}(d)) and may be a dominant factor of the resistance decrease.}
It is also worth noting that the film thickness showed only minor changes, in contrast to previous reports that attributed resistance reduction primarily to film thickening upon hydrogen absorption~\cite{das2020}. A plausible explanation is that the present films contain a large number of vacancies and exhibit discontinuous morphology in spite of its relatively large thickness: the hydrogenation establishes electrical path across the films in the lateral direction, thereby suppressing increase in the film thickness.
%A plausible explanation is that the present films contain a large number of vacancies, which annihilate during continuity improvement, thereby suppressing any net increase in the film thickness.

%\begin{figure}[t]
%    \centering
%    \includegraphics[width=1\linewidth]{figures/Rs/delta_Rs2.pdf}
%    \caption{(a) Working Ar pressure ($P_{\rm Ar}$) dependence of sheet resistance change ratio ($\Delta R_\mathrm{S}/R_0$). (b) $\Delta R_\mathrm{S}/R_0$ plotted against initial resistivity ($\rho_0$). We classified the data into two groups (see text).  }
%    \label{pardep}
%\end{figure}

%\begin{figure}[t]
%    \centering
%    \includegraphics[width=1\linewidth]{figures/XRD/xrd.pdf}
%    \caption{(a,b) XRD patterns of Pd(5.0 Pa) (a) and Pd(20 Pa) (b). Blue and red data represent the results obtained before and after the hydrogen exposure, respectively.}
%    \label{fig:xrd}
%\end{figure}

\begin{figure*}[t]
    \centering
    \includegraphics[width=1\linewidth]{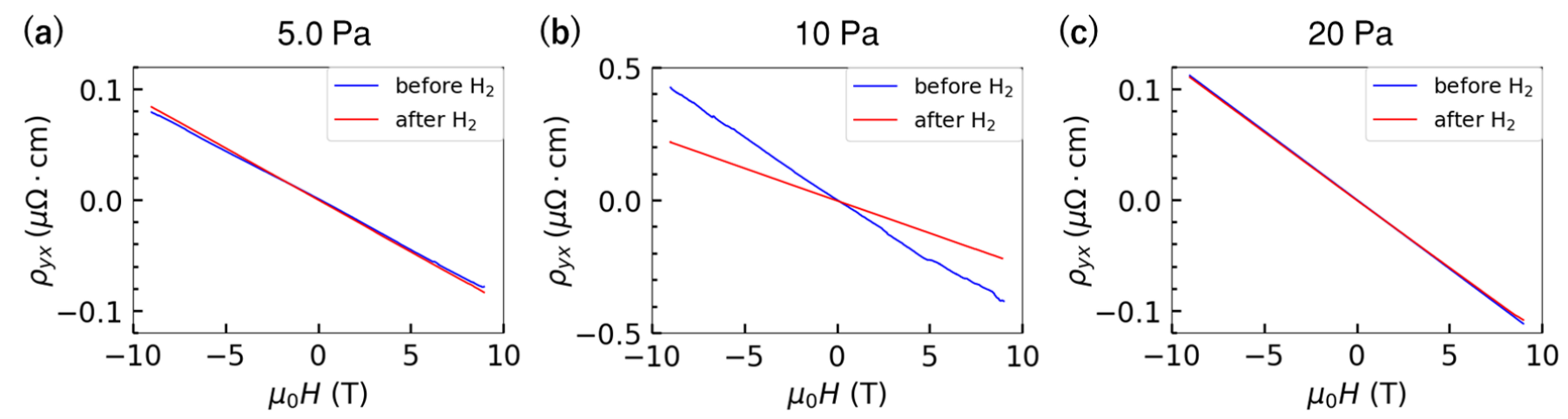}
    \caption{
    %Magnetic field ($H$) dependence of Hall resistivity ($\rho_{yx}$) for Pd(10 Pa). Blue and red data represent the results obtained before and after the hydrogen exposure, respectively.
    (a)-(c) Magnetic field ($H$) dependence of Hall resistivity ($\rho_{yx}$) for Pd(5.0 Pa) (a), Pd(10 Pa) (b) and Pd(20 Pa) (c). Blue and red data represent the results obtained before and after the hydrogen exposure, respectively.}
    \label{hall}
\end{figure*}

A closer inspection of Fig.~\ref{resistance}(d) reveals that the data for Pd(5.0 Pa) and Pd(10 Pa) deviate from the general trend: both films exhibited pronounced resistance changes despite relatively low initial resistances. It is notable that in these samples, the resistance decrease proceeds rather slowly as shown in Fig.~\ref{resistance}(b). 
In fact, the AFM images of Pd(5.0 Pa) before and after hydrogenation showed no significant increase in the grain size, as seen in Figs.~\ref{afm}(b) and (d). 
This observation suggests the presence of another mechanism that contributes to the decrease in the resistance. One plausible explanation is hydrogenation-induced crystallization of Pd. Generally, the amorphous phase shows a larger resistance than the crystalline phase for the same material. To examine this possibility, we compared the XRD patterns obtained before and after the hydrogen exposure, as shown in Fig.~\ref{afm} (f). The as-deposited Pd(5.0 Pa) film exhibited no distinct diffraction signal (i.e., it was nearly amorphous), whereas a clear Pd(111) peak emerged at $2\theta \sim 40^\circ$ after hydrogen absorption. Similar behavior was also observed for Pd(10 Pa): see Fig.~S1(c) in the supplementary material. These results indicate that the crystallization was promoted by hydrogenation.

%Another plausible explanation of the resistance decrease is hydrogenation-induced crystallization of Pd. Generally, the amorphous phase shows a larger resistance than the crystalline phase for the same material. To examine this possibility, we compared the XRD patterns obtained before and after the hydrogen exposure, as shown in Fig.~\ref{afm} (e) and (f): see Fig.~S1 in the supplementary material for the XRD patterns of Pd(2.5 Pa), Pd(7.5 Pa), Pd(10 Pa), and Pd(15 Pa). The as-deposited Pd(5.0 Pa) film exhibited no distinct diffraction signal (i.e., it was nearly amorphous), whereas a clear Pd(111) peak emerged at $2\theta \sim 40^\circ$ after hydrogen absorption. Similar behavior was also observed for Pd(10 Pa). Such a change of the XRD spectra indicates that the crystallization was promoted by hydrogenation. 

In contrast, for Pd(20 Pa), the diffraction peak of the Pd(111) orientation was observed both before and after hydrogen exposure as shown in Fig.~\ref{afm}(e). We confirmed that Pd(2.5 Pa) and Pd(15 Pa) also exhibited Pd(111) diffraction peaks prior to hydrogen exposure: see Fig.~S1(a) and (d) in the supplementary material. The difference in the initial film crystallinity supports the interpretation that hydrogenation-induced crystallization is the dominant mechanism for the decrease in resistance in Pd(5.0 Pa) and Pd(10 Pa). It is noteworthy that the formation of Pd hydride typically results in a peak shift toward lower angles~\cite{harumoto2017}; such a shift was not observed in all of our films, suggesting that hydride formation plays only a minor role. We remark that the reproducibility of these behaviors was verified across multiple Pd(5.0 Pa) and Pd(20 Pa). 

As a possible mechanism for the hydrogenation-induced crystallization of Pd films, the temperature increase due to Joule heating in the resistance measurements could accelerate the crystallization of Pd~\cite{xu1998}. However, this scenario is not applicable to the present case because the resistance change ratio is almost independent of the amplitude of the excitation currents: see Fig.~S2 in the supplementary material. 
As observed in the AFM images (Fig.~\ref{afm}), hydrogen exposure changed the grain structure, suggesting that the orientation of Pd may have improved during this process owing to its interaction with the substrate.
%During the process that grain structure changes with hydrogen exposure as observed in AFM images (Figs.~\ref{afm}(a)-(d)), the orientation of Pd may have improved owing to its interaction with the substrate.

%Within the present study, the underlying mechanisms for the hydrogenation-induced crystallizaion of Pd films remain uncovered. One previous study insisted that hydrogen exposure at high temperature can lead to the formation of vacancies, which might accelerate recrystallization of Pd~\cite{xu1998}. This scenario, however, might not be directly applicable in the present case since we did not warm up Pd films during the hydrogen exposure. Another study reported that ternary alloy nanoparticles containing Pd exhibited enhanced crystallinity upon hydrogen intercalation~\cite{cheng2023}, although its mechanism was not elucidated. Revealing the mechanisms of hydrogenation-induced crystallization in Pd thin films is an open question. 

We remark that in the case of Pd(7.5 Pa), a small diffraction peak appeared at $2\theta \sim 33^\circ$, which can be attributed to the (011) orientation of palladium oxide (PdO), both before and after hydrogenation: see Fig.~S1 (b) in the supplementary material. In contrast, the Pd (111) peak was absent prior to and following hydrogen exposure. These observations indicate that the significant resistance decrease cannot primarily be ascribed to the reduction of PdO into metallic Pd. Little XRD spectrum change and the rapid resistance change ($\sim 10$ min.) in Fig.~\ref{Champion} suggest that, in this film, resistance decrease is primarily governed by the improved electrical contact among the grains rather than crystallization. It might be possible that PdO obstructing the electrical paths among Pd grains enhances the electrical discontinuity of the film, which can in turn result in a pronounced reduction in film resistance once continuous conduction paths are formed through hydrogenation. We also note that the emergence of the initial resistance increase (Fig.~\ref{Champion}) is linked with the presence of the PdO phase, as revealed by the investigation for a Pd film thermally annealed in air: see Sec.~III of the supplementary material.
%\textcolor{red}{We also note that in the case of Pd(2.5 Pa), the Pd(111) peak was enhanced by the hydrogen exposure while the resistance change ratio was relatively smaller than that of Pd(5.0 Pa) and Pd(10 Pa). We consider that the improvement of the crystallinity may also be responsible for the resistance decrease in Pd(2.5 Pa). Since the amorphous phase generally exhibits higher resistance than crystalline phase, the resistance change ratio is expected to be larger when an initially amorphous film undergoes crystallization than when an already crystalline film becomes further crystallized.}

Previous studies have suggested that the density of states at the Fermi level can be modulated by hydrogenation~\cite{nemirovich2021,eberhardt1983}. Hence, we examined the modulation of carrier density upon hydrogen absorption via the Hall-effect measurement. 
Figure~\ref{hall} shows the magnetic-field dependence of the Hall resistivity for Pd(5.0 Pa), Pd(10 Pa), and Pd(20 Pa), measured at room temperature using a Physical Property Measurement System (PPMS, Quantum Design). 
The slopes, i.e. the Hall coefficients, are negative, meaning that the dominant carriers are electrons. The electron density is on the order of $10^{22}$--$10^{23}$ cm$^{-3}$, which is consistent with typical metallic behavior.
Pd(5.0 Pa) and Pd(20 Pa) showed little change of the electron density by hydrogen exposure, while the electron density of Pd(10 Pa) was doubled after the hydrogen absorption. Even though the increased electron density of Pd(10 Pa) likely contributed to the observed decrease in the resistance, the resistance change, which was apparently exceeding 50 \%, cannot be explained solely by this carrier density modulation. The decrease in resistance arises from a combination of other factors, including the improved electrical contact and hydrogenation-induced crystallization.
%Previous studies have suggested that the density of states at the Fermi level can be modulated by hydrogenation~\cite{nemirovich2021,eberhardt1983}. Hence, we examined the modulation of carrier density upon hydrogen absorption via the Hall-effect measurement. Figure~\ref{hall} shows the magnetic-field dependence of the Hall resistivity for Pd(10 Pa), measured at room temperature using a Physical Property Measurement System (PPMS, Quantum Design). The slope, i.e. the Hall coefficient, is negative, meaning that the dominant carriers are electrons. Prior to hydrogen exposure, the electron density was estimated to be $1.4 \times 10^{22}$ cm$^{-3}$, which is consistent with typical metallic behavior. After hydrogen absorption, the electron density approximately doubled, which likely contributed to the observed decrease in the resistance. However, the resistance change is apparently larger than 50 \%, and thus cannot be explained solely by this carrier density modulation. Rather, it arises from a combination of other factors, including the improved electrical contact and hydrogenation-induced crystallization.

The carrier increase of Pd(10 Pa) is attributed to electron doping by the absorbed hydrogens~\cite{mackliet1966}. Although the modulation of the band structure and density of states~\cite{nemirovich2021,eberhardt1983} at the Fermi level is another plausible origin, little change in the carrier density in Pd(5.0 Pa), where the crystallinity is significantly improved after hydrogen exposure, suggests that the change in the electronic structure is minor. We remark that the increase in carrier density of Pd by several-fold upon hydrogenation is feasible according to a previous paper~\cite{wisniewski1971}.
%Although the predominant cause of the carrier increase of Pd(10 Pa) was not identified within the scope of this work, electron doping by the absorbed hydrogens~\cite{mackliet1966} and/or the modulation of the electron density of states~\cite{nemirovich2021,eberhardt1983} might be responsible for it. Note that a previous study reported that the carrier density of palladium can increase by several-fold upon hydrogenation at high hydrogen concentrations~\cite{wisniewski1971}.

In summary, we observed a significant decrease in the electrical resistance of the Pd thin films simply deposited under a high working Ar pressure using a compact film coater. The resistance reduction ratio reached up to a factor of $1/335$, being predominant among those reported previously. Characterization of the films including atomic force microscopy observation and X-ray diffraction measurement suggested two primary mechanisms responsible for such significant resistance reduction: (i) the enhancement of electrical contacts among Pd grains, and (ii) hydrogenation-induced crystallization of Pd. We found that the former mechanism induced a rapid decrease in resistance, with the change ratio scaling with the initial resistivity, whereas the latter proceeded more gradually and did not exhibit such a correlation. These results provide a simple yet effective strategy for improving hydrogen sensor performance and contribute to a more comprehensive understanding of the hydrogenation process in Pd.

\vspace{10pt}
See the supplementary material for the other XRD data, current dependence of resistance change ratio, and results for thermally annealed Pd.

\section*{Acknowledgments}
The authors thank Dr. H. Okuma and Prof. K. Ueno for support in the XRD measurements.
This work was supported by Tanaka Kikinzoku Memorial Foundation, by JST FOREST Program, Grant Number JPMJFR203H, and by JSPS KAKENHI (Grant Nos. JP24K23028, JP25H01613, JP25H02122, JP25K17907, JP23K26525, JP24H01177, JP24K21726, JP24K00566, and JP25H00611).
\section*{Author declarations}
\subsection*{Conflict of Interest}
The authors declare that they have no conflicts of interest.
\subsection*{Author Contributions}
Y.S. planned the study in discussions with T.K., and Y.I. and T.K. performed the experiments with support from Y.S. All the authors participated in the discussion of the experimental results. Y.I. and T.K. wrote the manuscript with input from Y.S. 
\section*{Data Availability}
All the data needed to reach the conclusion of this study are available in the article. Additional data related to this paper are available from the corresponding author upon reasonable request.

\section*{References}
\bibliography{ref}

\end{document}